# Paired Orbitals for Different Spins equations

Igor Zilberberg*
*Boreskov Institute of Catalysis and Novosibirsk State University, Novosibirsk 630090, Russian Federation*
Sergey Ph. Ruzankin
*Boreskov Institute of Catalysis, Novosibirsk 630090, Russian Federation*

Abstract

Eigenvalue-type equations for Löwdin-Amos-Hall spin-paired (corresponding) orbitals are developed to provide an alternative to the standard spin-polarized Hartree-Fock or Kohn-Sham equations. Obtained equations are non-canonical unrestricted Hartree-Fock-type equations in which non-canonical orbitals are fixed to be biorthogonal spin-paired orbitals. To derive paired orbitals for different spins (PODS) equations there has been applied Adams-Gilbert "localizing" operator approach. PODS equations are especially useful for treatment of the broken-symmetry solutions for antiferromagnetic materials.

## I. INTRODUCTION

Different orbitals for different spins (DODS) approximation – introduced by Löwdin[1] as a straightforward approach to account for the Coulomb correlation in the Hartree-Fock theory (HFT) – is now widely used to treat large open-shell molecular systems mostly in the framework of density functional theory (DFT).[2]

Single determinant wave function in the DODS approximation is built from independently varied orbitals for either spin:

$$\Psi^{DODS} = (N!)^{-1/2} \det\{\psi_1^\alpha \alpha ... \psi_{N_\alpha}^\alpha \alpha \psi_1^\beta \beta ... \psi_{N_\beta}^\beta \beta\} \qquad (1)$$

where spin-polarized orbitals $\left\{ \left|\psi_i^\alpha\right\rangle, i = 1, N_{basis} \right\}$ and $\left\{ \left|\psi_i^\beta\right\rangle, i = 1, N_{basis} \right\}$ are solutions of the unrestricted Hartree-Fock (UHF) equations or Kohn-Sham (UKS) equations:

* igor@catalysis.ru

$$\begin{cases} F^{\alpha}\left|\psi_i^{\alpha}\right\rangle = \varepsilon_i^{\alpha}\left|\psi_i^{\alpha}\right\rangle \\ F^{\beta}\left|\psi_i^{\beta}\right\rangle = \varepsilon_i^{\beta}\left|\psi_i^{\beta}\right\rangle \end{cases} \quad (2)$$

Here $N_{basis}$ is a number of basis functions, $N_\alpha$ and $N_\beta$ are the numbers of the α and β electrons, respectively. For simplicity there is supposed that $N_{basis} >> N_\alpha \geq N_\beta$. The α and β orbital sets are mutually nonorthogonal.

These equations are referred to as canonical to point out that the $\left|\psi_i^{\alpha}\right\rangle$ and $\left|\psi_i^{\beta}\right\rangle$ orbitals are eigenvectors of the one-electron operators $F^{\alpha}$ and $F^{\beta}$, respectively. As well known, unitary transformations of the occupied orbitals

$$\begin{aligned} \left|a_i\right\rangle &= \sum_k^{N_\alpha}\left|\psi_k^{\alpha}\right\rangle U_{ki}, i = 1, N_\alpha \\ \left|b_j\right\rangle &= \sum_k^{N_\beta}\left|\psi_k^{\beta}\right\rangle V_{kj}, j = 1, N_\beta \end{aligned} \quad (3)$$

(where $\mathbf{U}^{\dagger}\mathbf{U}=\mathbf{U}\mathbf{U}^{\dagger}=\mathbf{I}_\alpha$ ($N_\alpha$ by $N_\alpha$ identity matrix) and $\mathbf{V}^{\dagger}\mathbf{V}=\mathbf{V}\mathbf{V}^{\dagger}=\mathbf{I}_\beta$ ($N_\beta$ by $N_\beta$ identity matrix)) leave determinant (1) unchanged within a phase factor. The spin density operators $\rho^{\alpha} = \sum_k^{N_\alpha}\left|\psi_k^{\alpha}\right\rangle\left\langle\psi_k^{\alpha}\right|$ and $\rho^{\beta} = \sum_k^{N_\beta}\left|\psi_k^{\beta}\right\rangle\left\langle\psi_k^{\beta}\right|$, and the total energy are invariant with respect to such transformations. This invariance allows using other (non-canonical) sets of orbitals to get a specific insight into the electron structure of particular system. Corresponding non-canonical equations are:

$$\begin{cases} F^{\alpha}\left|a_i\right\rangle = \sum_j^{N_\alpha}\left|a_j\right\rangle \varepsilon_{ji}^{\alpha} \\ F^{\beta}\left|b_k\right\rangle = \sum_l^{N_\beta}\left|b_l\right\rangle \varepsilon_{lk}^{\beta} \end{cases} \quad (4)$$

One of the popular non-canonical sets is the set of paired (or corresponding) orbitals introduced by Löwdin[3], Amos and Hall[4] and Karadakov[5]. A simple proof of the pairing theorem was given by Mayer.[6] In practice "pairing" transformation matrices **U** and **V** can be obtained by singular value decomposition (SVD) of the rectangular matrix[7] of overlap between the α and β orbital sets of the standard (canonical) unrestricted solution.

By construction paired occupied orbital sets are biorthogonal:[4]

$$\begin{aligned} \langle a_i | a_{i'} \rangle &= \delta_{ii'} \\ \langle b_j | b_{j'} \rangle &= \delta_{jj'} \\ \langle b_j | a_i \rangle &= \delta_{ji} t_j \end{aligned} \qquad (5)$$

where the overlap is usually defined as $t_j = \cos 2\theta_j$. Paired virtual orbitals satisfy analogous orthogonality conditions:[8]

$$\begin{aligned} \langle a_j^v | a_{j'}^v \rangle &= \delta_{jj'} \\ \langle b_j^v | a_{j'} \rangle = \langle a_j^v | b_{j'} \rangle &= \delta_{jj'} \sin 2\theta_j = \delta_{jj'} (1 - t_j^2)^{1/2} \\ \langle b_j^v | a_{j'}^v \rangle &= -\delta_{jj'} t_j \end{aligned} \qquad (6)$$

From the above given conditions it is clear that any β-spin orbital is completely determined by the pair of corresponding α occupied and α unoccupied orbitals:[8]

$$|b_j\rangle = |a_j\rangle \cos 2\theta_j + |a_j^v\rangle \sin 2\theta_j \qquad (7)$$

where $j = 1, N_\beta$ and $i = 1, N_\alpha$. The expansion of $|a_i\rangle$ over $|b_i\rangle$ and $|b_i^v\rangle$ is analogous.

In case of the system with antiferromagnetically coupled magnetic moments modeled within the broken-symmetry approach developed by Noodleman[9] paired orbitals might be considered as orbital channels providing such coupling. In the basis of paired orbitals spin contamination defined as $\langle S^2 - S_z(S_z+1) \rangle$ (where $S_z = (N_\alpha - N_\beta)/2$) can be interpreted as the effective number of pairs of spatially separated orbitals occupied by two electrons with anti-parallel spins. In addition, in the basis of paired orbitals unrestricted determinant can be expanded in a linear combination of restricted determinants allowing one to assign a structure in question in terms of idealized covalent and charge-transfer configurations.[8] Using this approach called “S2-expansion technique” there has been investigated a few open-shell systems such as the Fe(II)-NO complex[10], the Fe(II)-nitrobenzene complex[11], ruthenium complexes with redox-active quinonoid ligands[12] and the Fe(II)-porphyrin nitroxyl complexes[13].

In some applications it is desirable to control orbital magnetic channels at each iteration step toward self-consistency since their parameters $t_i$ (overlap between paired orbitals) might be considered as the order parameters in the transition between symmetric non-magnetic $(t_i = 1)$ phase and asymmetric anti-ferromagnetic $(0 \le t_i < 1)$ phase. For these purposes a set of equations for paired orbitals would be the most natural tool. Developing of such equations is the aim of the present paper. These equations, called below paired orbitals for different spins (PODS) equations, are derived as non-canonical equations with an additional requirement for the α and β orbital sets to be biorthogonal. To develop PODS equations in the form of eigenvalue equations there has been used an operator of Gilbert's-type

$$R = F + \rho A \rho + (1-\rho)B(1-\rho) \qquad (8)$$

(where $\rho$ is the Dirac-Fock density, $A$ and $B$ are arbitrary Hermitian operators) eigenvectors of which are non-canonical HF orbitals[14]. In the basis of the HF orbitals the matrix of modified Fock operator (8) is block diagonal since operator $\rho A \rho$ (and $(1-\rho)B(1-\rho)$ as well) doesn't couple occupied and unoccupied orbitals. The block of occupied orbitals and unoccupied orbitals become non-diagonal as operator $\rho A \rho$ and $(1-\rho)B(1-\rho)$ mixes occupied orbitals and unoccupied orbitals, respectively, between themselves. This mixing doesn't affect density and total energy though allowing one to fix a particular non-canonical orbital set, e.g. that in which orbitals are localized on atomic centers in molecule. As was pointed out by Gilbert[14], "non-canonicalization" of both the occupied and unoccupied orbitals is determined by the operators $A$ and $B$ chosen to obtain a desired linear combination of the standard occupied and virtual orbitals. In the present work "non-canonicalization" operators defined separately for spin-polarized Fock (or Kohn-Sham) operators are constructed in such a way to diagonalize the overlap between the α and β orbitals. In the paper two possible choices of such operators are presented along with corresponding sets of equations.

In section II the PODS equations are derived in the form Adams-Gilbert-type effective operators for either spins constructed from spin densities. The eigenvalues of these operators are in fact squared overlaps between paired α and β orbitals. In section III an alternative form of the PODS equations is developed using Edmiston-Ruedenberg-like “non-canonicalization” operators. Their eigenvalues can be considered as the HF (KS) orbital energies split by the field of “pairing” operator.

## II. PAIRED ORBITAL ADAMS-GILBERT-LIKE EQUATIONS

Non-canonical unrestricted equations (4) are easily transformed into pseudo-eigenvalue equations by means of Adams-Gilbert approach to construct the effective operators $R^{\alpha}$ and $R^{\beta}$ for either spin:

$$R^{\sigma} = F^{\sigma} + \rho^{\sigma}(\Omega^{\sigma} - F^{\sigma})\rho^{\sigma} + (1-\rho^{\sigma})(\Lambda^{\sigma} - F^{\sigma})(1-\rho^{\sigma})$$

$$= (1-\rho^{\sigma})F^{\sigma}\rho^{\sigma} + \rho^{\sigma}F^{\sigma}(1-\rho^{\sigma}) + \rho^{\sigma}\Omega^{\sigma}\rho^{\sigma} + (1-\rho^{\sigma})\Lambda^{\sigma}(1-\rho^{\sigma}) \qquad (9)$$

where σ stands for the α or β spin and $\Omega^{\sigma}$ and $\Lambda^{\sigma}$ are arbitrary Hermitian operators. Since $F^{\sigma}, \rho^{\sigma}, \Omega^{\sigma}$ and $\Lambda^{\sigma}$ are Hermitian operators, effective operator $R^{\sigma}$ is Hermitian as well. Non-canonical Hartree-Fock-type orbitals are eigenvectors of these operators:

$$\begin{cases} R^{\alpha}\left|a_i\right\rangle = \eta_i^{\alpha}\left|a_i\right\rangle \\ R^{\beta}\left|b_j\right\rangle = \eta_j^{\beta}\left|b_j\right\rangle \end{cases} \qquad (10)$$

Fixation of paired α and β orbitals from all possible non-canonical sets should result from additional constraint for the integrals of overlap between α and β orbitals to form a diagonal matrix which is achieved by means of suitably defined operators $\Omega^{\alpha}, \Lambda^{\alpha}$ and $\Omega^{\beta}, \Lambda^{\beta}$. These operators are suggested in the present work to be constructed from the $\rho^{\alpha}$ and $\rho^{\beta}$ spin densities since the density matrix elements in the basis of unrestricted α and β orbitals are determined by the overlap between these orbitals (see Appendix):

$$\Omega^{\alpha} = \rho^{\alpha} + \rho^{\beta} = \rho \qquad (11)$$

$$\Lambda^{\alpha} = \rho^{\beta} \tag{12}$$

and

$$\Omega^{\beta} = \rho^{\alpha} + \rho^{\beta} = \rho \tag{13}$$

$$\Lambda^{\beta} = \rho^{\alpha}\,. \tag{14}$$

Correspondingly, the effective operators become:

$$R^{\alpha} = (1-\rho^{\alpha})F^{\alpha}\rho^{\alpha} + \rho^{\alpha}F^{\alpha}(1-\rho^{\alpha}) + \rho^{\alpha}\rho\rho^{\alpha} + (1-\rho^{\alpha})\rho^{\beta}(1-\rho^{\alpha}) \tag{15}$$

$$R^{\beta} = (1-\rho^{\beta})F^{\beta}\rho^{\beta} + \rho^{\beta}F^{\beta}(1-\rho^{\beta}) + \rho^{\beta}\rho\rho^{\beta} + (1-\rho^{\beta})\rho^{\alpha}(1-\rho^{\beta}) \tag{16}$$

In the basis of occupied $\left|a\right\rangle$ and unoccupied $\left|a^{v}\right\rangle$ orbitals the $R^{\alpha}$ operator matrix has a simple block structure:

$$\mathbf{R}^{\alpha} = \left[\begin{array}{cc|cc} 1+\left\langle a_i\middle|\rho^{\beta}\middle|a_i\right\rangle & \left\langle a_i\middle|\rho^{\beta}\middle|a_j\right\rangle & \multicolumn{2}{c}{\multirow{2}{*}{$\left\langle a\middle|F^{\alpha}\middle|a^{v}\right\rangle$}} \\ \left\langle a_j\middle|\rho^{\beta}\middle|a_i\right\rangle & 1+\left\langle a_j\middle|\rho^{\beta}\middle|a_j\right\rangle & & \\ \hline \multicolumn{2}{c|}{\multirow{2}{*}{$\left\langle a^{v}\middle|F^{\alpha}\middle|a\right\rangle$}} & \left\langle a_i^{v}\middle|\rho^{\beta}\middle|a_i^{v}\right\rangle & \left\langle a_i^{v}\middle|\rho^{\beta}\middle|a_j^{v}\right\rangle \\ & & \left\langle a_j^{v}\middle|\rho^{\beta}\middle|a_i^{v}\right\rangle & \left\langle a_j^{v}\middle|\rho^{\beta}\middle|a_j^{v}\right\rangle \end{array}\right] \tag{17}$$

The $\mathbf{R}^{\beta}$ matrix has a similar structure. Both matrices are Hermitian and can be always diagonalized by suitable unitary transformations. Non-diagonal blocks disappear after diagonalization providing the condition of the $\left|a_i\right\rangle$ and $\left|b_j\right\rangle$ to be Hartree-Fock (Kohn-Sham) orbitals, i.e. the orbitals which mix between themselves separately (withing the occupied and unoccupied manifolds) upon the action of the $F^{\alpha}$ and $F^{\beta}$ operators. Diagonal blocks are responsible for pairing of the α and β orbitals. Once the $\left\langle a_j\middle|\rho^{\beta}\middle|a_i\right\rangle$ matrix is diagonalized, the $\left|a_i\right\rangle$ and $\left|b_j\right\rangle$ sets become paired, i.e. $\left\langle a_i\middle|b_j\right\rangle = \delta_{ij}t_j$ for $j = 1, N_{\beta}$. Diagonalizing of the $\left\langle a_l^{v}\middle|\rho^{\beta}\middle|a_k^{v}\right\rangle$ matrix is equivalent to pairing of the unoccupied orbitals (see eq. 6). The same is true for the β orbitals.

Eigenvalues of the $\mathbf{R}^{\alpha}$ and $\mathbf{R}^{\beta}$ matrices are determined by the overlap in the αβ pairs $t_i = \langle a_i | b_i \rangle$ (see eq. 5-7):

$$\begin{aligned}
&\begin{cases} \eta_i^{\alpha} = \eta_i^{\beta} = 1 + t_i^2 \\ \eta_{i+N_\alpha}^{\alpha} = \eta_{i+N_\beta}^{\beta} = 1 - t_i^2 \end{cases} && i = 1, N_\beta \\
&\eta_i^{\alpha} = 1 && i = N_\beta + 1, N_\alpha \\
&\eta_k^{\alpha} = 0 && k = N_\alpha + N_\beta + 1, N_{basis} \\
&\eta_k^{\beta} = 0 && k = 2N_\beta + 1, N_{basis}
\end{aligned} \tag{18}$$

If the number of α electrons is larger than the number of β electrons (i.e. the system has a nonzero magnetic moment) then there is a set of eigenvalues $\eta_i^{\alpha} (i = N_\beta + 1, N_\alpha)$ each being equal exactly 1. This set of truly unpaired α electrons makes the only difference between the α and β eigenvalues. Degenerate unoccupied α and β orbitals are also unpaired having zero eigenvalues. To achieve self-consistency it is important to distinguish between all these groups of orbitals at each iteration since the α and β densities (responsible for the total energy) are constructed only from occupied orbitals. In practice occupied and unoccupied eigenvectors are ordered by their eigenvalues and so some unwanted mixing may happen, e.g. for intermediate values of squared overlap close to 0.5. Such difficulties can be avoided by the use of shifting operators. In fact, above-defined operator $\Omega^{\alpha}$ already contains such operator $\rho^{\alpha}$ which shifts all occupied orbitals by 1 atomic unit up keeping the vectors (and so the densities and total energy) unchanged.

## III. EDMISTON-RUEDENBERG-LIKE EQUATIONS

Alternative form of paired-orbital equations can be obtained within Edmiston-Ruedenberg approach for localizing orbitals[15]. In this approach to obtain localized orbitals the off-diagonal elements of the Fock matrix are substituted by the matrix elements of some non-local operator. This allows one to fix a particular non-canonical HF solution from all possible solutions. Applying this idea for paired-orbital equations one can obtain the following matrix of the $R^{\alpha}$ operator ($R^{\beta}$ is constructed in analogous way):

$$
\mathbf{R}^{\alpha}=\left[\begin{array}{cc|cc}
\langle a_i|F^{\alpha}|a_i\rangle & \langle a_i|\rho^{\beta}|a_j\rangle & \multicolumn{2}{c}{\langle a|F^{\alpha}|a^{v}\rangle} \\
\langle a_j|\rho^{\beta}|a_i\rangle & \langle a_j|F^{\alpha}|a_j\rangle & & \\
\hline
\multicolumn{2}{c|}{\langle a^{v}|F^{\alpha}|a\rangle} & \langle a_i^{v}|F^{\alpha}|a_i^{v}\rangle & \langle a_i^{v}|\rho^{\beta}|a_j^{v}\rangle \\
 & & \langle a_j^{v}|\rho^{\beta}|a_i^{v}\rangle & \langle a_j^{v}|F^{\alpha}|a_j^{v}\rangle
\end{array}\right]
$$

Corresponding operator is defined as follows

$$
R^{\alpha}=(1-\rho^{\alpha})F^{\alpha}\rho^{\alpha}+\rho^{\alpha}F^{\alpha}(1-\rho^{\alpha})+\sum_{i}^{occ}|a_i\rangle\langle a_i|F^{\alpha}|a_i\rangle\langle a_i|+\sum_{k}^{uno}|a_k^{v}\rangle\langle a_k^{v}|F^{\alpha}|a_k^{v}\rangle\langle a_k^{v}|+
$$
$$
\rho^{\alpha}\rho^{\beta}\rho^{\alpha}-\sum_{i}^{occ}|a_i\rangle\langle a_i|\rho^{\beta}|a_i\rangle\langle a_i|+(1-\rho^{\alpha})\rho^{\beta}(1-\rho^{\alpha})-\sum_{k}^{uno}|a_k^{v}\rangle\langle a_k^{v}|\rho^{\beta}|a_k^{v}\rangle\langle a_k^{v}|, \qquad (19)
$$

where *occ* and *uno* are the numbers of occupied and unoccupied (virtual) orbitals respectively. Eigenvalues of these operators are interpreted as follows. Let us suppose that for some system there is known a canonical UHF (or UKS) solution. Then diagonal matrix elements of the $R^{\alpha}$ and $R^{\beta}$ operators in the basis of these canonical orbitals will be the energies ($\varepsilon_i^{\alpha}$ and $\varepsilon_i^{\beta}$) of these orbitals. In accordance with a general property of square matrices the trace of matrix is the sum of its eigenvalues.

$$
\sum_{i}^{N_{basis}}\eta_i^{\sigma}=\sum_{i}^{N_{basis}}\varepsilon_i^{\sigma}
$$
$$
(\sigma=\alpha,\beta)
$$

Since the only nonzero off-diagonal elements of the $\mathbf{R}^{\alpha}$ and $\mathbf{R}^{\beta}$ matrices in the canonical basis are $\langle a_j|\rho^{\beta}|a_i\rangle$ and $\langle b_j|\rho^{\alpha}|b_i\rangle$ these matrices are block diagonal. Therefore, for each block (occupied and unoccupied ones) analogous sum rule holds:

$$
\sum_{i}^{N_{\alpha}}\eta_i^{\alpha}=\sum_{i}^{N_{\alpha}}\varepsilon_i^{a} \qquad \text{for occupied orbitals and}
$$

$$
\sum_{i=N_{\alpha}+1}^{N_{basis}}\eta_i^{\alpha}=\sum_{i=N_{\alpha}+1}^{N_{basis}}\varepsilon_i^{a} \qquad \text{for unoccupied orbitals.}
$$

An analogous rule holds for the β eigenvalues $\eta^{\beta}$ as well.

Therefore the eigenvalues of the effective "pairing" operators can be interpreted as the standard α and β one-electron energies split by a field (proportional to the density of the β and α electrons, respectively) such that the center of gravity of the one-electron levels is unchanged.

## IV. CONCLUDING REMARKS

In this work there have been obtained modified (so-called PODS) unrestricted equations equally applicable for both Hartree-Fock and Kohn-Sham theory. Unlike standard unrestricted solutions, PODS orbitals for either spin are all biorthogonal (paired). Another feature of the PODS equations is that the eigenvalues of paired α and β orbitals are equal. The PODS operators are not uniquely defined and so are their eigenvalues. So far two versions of PODS equations have been developed though it is unclear yet which of two versions is more practical from computational point of view.

It is worthwhile to note that a spin-broken-symmetry $S_z = 0$ solution is not produced automatically within the PODS approach as well as in the standard unrestricted theory due to the equivalence of the α and β orbital sets in the initial guess. In such cases some known recipes of the initial guess "brokenization" have to be applied.

## APPENDIX

Consider matrix elements of the $\rho^{\beta}$ ($\rho^{\alpha}$) operator in the basis of α(β) occupied orbitals appeared in the unrestricted Hartree-Fock or Kohn-Sham methods.

$$\left\langle \psi_i^{\alpha} \middle| \rho^{\beta} \middle| \psi_j^{\alpha} \right\rangle = \sum_{k=1}^{N_\beta} \left\langle \psi_i^{\alpha} \middle| \psi_k^{\beta} \right\rangle \left\langle \psi_k^{\beta} \middle| \psi_j^{\alpha} \right\rangle = (\mathbf{O}^{\dagger}\mathbf{O})_{ij} \qquad (\text{for } i, j = 1, N_\alpha) \qquad \text{(A1)}$$

where $\mathbf{O}_{ki} = \left\langle \psi_k^{\beta} \middle| \psi_i^{\alpha} \right\rangle$. These integrals compose a $N_\beta \times N_\alpha$ rectangular matrix **O**. The $\mathbf{O}^{\dagger}\mathbf{O}$ matrix is a $N_\alpha \times N_\alpha$ Hermitian and positive definite matrix with a rank $\le N_\beta$.[4,7] This matrix can be always diagonalized by a unitary matrix $\mathbf{U}$ :

$\mathbf{O}^{\dagger}\mathbf{O} = \mathbf{U}\mathbf{D}_{\boldsymbol{\alpha}}\mathbf{U}^{\dagger}$, (A2)

where

$\mathbf{D}_{\alpha} = diag(d_1,...,d_{N_\beta},0,...,0)$ (A3)

is a $N_\alpha \times N_\alpha$ diagonal matrix having at minimum $(N_\alpha - N_\beta)$ zero eigenvalues.[4,7]

Matrix elements of the $\rho^\alpha$ operator in the basis of occupied β orbitals are expressed in the analogous way

$$\left\langle \psi_i^\beta \middle| \rho^\alpha \middle| \psi_j^\beta \right\rangle = \sum_k^{N_\alpha} \left\langle \psi_i^\beta \middle| \psi_k^\alpha \right\rangle \left\langle \psi_k^\alpha \middle| \psi_j^\beta \right\rangle = (\mathbf{O}\mathbf{O}^{\dagger})_{ij} \qquad (\text{for } i,j = 1, N_\beta) \qquad \text{(A4)}$$

Since the $N_\beta \times N_\beta$ matrix $\mathbf{O}\mathbf{O}^{\dagger}$ is Hermitian, then there exists a unitary matrix $\mathbf{V}$ diagonalizing $\mathbf{O}\mathbf{O}^{\dagger}$:

$\mathbf{O}\mathbf{O}^{\dagger} = \mathbf{V}\mathbf{D}_{\boldsymbol{\beta}}\mathbf{V}^{\dagger}$ (A5)

Eigenvalues of the $\mathbf{O}\mathbf{O}^{\dagger}$ matrix composing a $N_\beta \times N_\beta$ diagonal matrix

$\mathbf{D}_\beta = diag(d_1,...,d_{N_\beta})$ coincide with non-zero eigenvalues of the $\mathbf{O}^{\dagger}\mathbf{O}$ matrix.[4,7]

Therefore, transformed orbitals

$$\left| a_i \right\rangle = \sum_k^{N_\alpha} \left| \psi_k^\alpha \right\rangle U_{ki}, \; i = 1, N_\alpha \qquad \text{(A6)}$$

and

$$\left| b_j \right\rangle = \sum_k^{N_\beta} \left| \psi_k^\beta \right\rangle V_{kj}, \; j = 1, N_\beta \qquad \text{(A7)}$$

are paired (see eq. 6):

$\left\langle b_j \middle| a_i \right\rangle = \delta_{ji} t_j$ for $i = 1, N_\alpha$ and $j = 1, N_\beta$. (A8)

where overlaps are chosen to be non-negative

$t_i = d_i^{1/2} (i = 1, N_\beta)$. (A9)

Matrix elements of the $\rho^{\alpha}$ and $\rho^{\beta}$ operator in the basis of the β unoccupied orbitals and α unoccupied orbitals, respectively, are analogously diagonalized (see eq. 6).